\documentclass[aps,twocolumn,showpacs,amsmath,amssymb,pre,superscriptaddress, floatfix]{revtex4-1}

\usepackage{graphicx}
\usepackage{dcolumn}
\usepackage{bm}
\usepackage{amssymb}
\usepackage{multirow}

\usepackage{xr}
\newcommand{\1}{\begin{equation}}
\newcommand{\2}{\end{equation}}
\newcommand{\ea}{\begin{eqnarray}} 
\newcommand{\ee}{\end{eqnarray}}

\newcommand{\I}{{ {\rm i}  }}

\begin{document}

\title{Pattern Formation in Chemically Interacting Active Rotors with Self-Propulsion}

\date{\today}

\author{Benno Liebchen}\email[]{Benno.Liebchen@staffmail.ed.ac.uk}\affiliation{SUPA, School of Physics and Astronomy, University of Edinburgh, Edinburgh EH9 3FD, United Kingdom}%
\author {Michael E. Cates}\affiliation{DAMTP, Centre for Mathematical Sciences, University of Cambridge, Cambridge CB3 0WA, United Kingdom}
\author{Davide Marenduzzo}\affiliation{SUPA, School of Physics and Astronomy, University of Edinburgh, Edinburgh EH9 3FD, United Kingdom}

\begin{abstract}

We demonstrate that active rotations in chemically signalling particles, such as autochemotactic {\it E. coli} close to walls, create a route for
 pattern formation based on a nonlinear yet deterministic instability mechanism.
For slow rotations, we find a transient persistence of the uniform state, followed by a sudden formation of clusters 
contingent on locking of the average propulsion direction by chemotaxis.
These clusters coarsen, which results in phase separation into a dense and a dilute region.
Faster rotations arrest phase separation leading to a global travelling wave of rotors with synchronized roto-translational motion.
Our results elucidate the physics resulting from the competition of two generic paradigms in active matter, chemotaxis and active rotations, and show that the latter provides a tool 
to design programmable self-assembly of active matter, for example to control coarsening. 
\end{abstract}
\maketitle

\begin{figure*}
 \begin{center}
  \includegraphics[width=\textwidth]{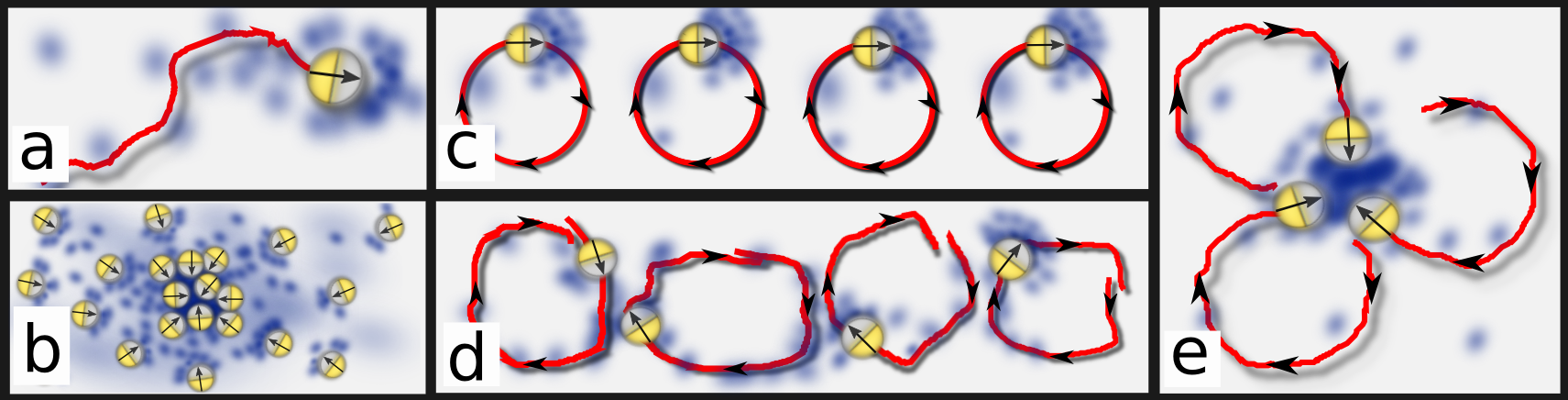}
\caption{\small Cartoons showing the basic setup and the locking instability (see text). 
a.) Chemotactic colloid/bacterium swimming up a self-produced chemical gradient.
b.) Keller-Segel instability. 
c.) Uniform phase of coherent rotors.
d.) Rotors leave chemical trail dephasing coherent rotations. 
e.) Possible seed configuration for the locking instability.}
\label{fig1}
 \end{center}
\end{figure*}

Various self-propelled microorganisms, like the bacterium {\it E. coli}~\cite{Budrene1991,Budrene1995,Mittal2003} or the slime mold {\it Dictyostelium disoideum} (Dicty)~\cite{Gerisch1982} 
can adapt their motion in response to gradients of chemicals which they produce themselves
(Fig.~\ref{fig1}a). 
Such a chemical `signalling' can induce spontaneous aggregation through a positive feedback (Fig.~\ref{fig1}b): 
an initial positive chemical density fluctuation attracts microorganisms, which produce further chemicals;
this amplifies the initial fluctuation supporting the recruitment of further microoganisms, which eventually condense into a 
macroscopic cluster coexisting with a dilute microbial bath (Fig.~\ref{fig1}b). 
Quantified by the Keller-Segel (KS) model in the 1970s \cite{Keller1970,Keller1971}, this signalling-route to phase separation 
now serves as a prototypical example for self-organization in nonequilibrium~\cite{Haken1984}. It is crucial for the life cycle of Dicty \cite{Gerisch1982}; 
it is also one possible mechanism to explain pattern formation in bacterial colonies within agar~\cite{Budrene1991,Budrene1995}, alongside an alternative 
explanation based on arrested motility induced phase separation \cite{Cates2010}.

Currently, the KS instability is attracting renewed attention in active synthetic Janus colloids. 
These particles catalyse reactions within a chemical bath, yielding gradients which drive self-propulsion via diffusiophoresis or a similar mechanism. Importantly, 
other colloids may feel rotational torques caused by the same long-ranged gradients and respond to them by adapting their swimming direction, thereby providing a 
synthetic analog of chemotactic signalling~\cite{Hong2007,Theurkauff2012,Saha2014}. The same KS equations can therefore be transferred from the microbiological 
world to phoretic colloids~\cite{Theurkauff2012,Taktikos2012,Meyer2014,Pohl2014}. Here, either the KS instability, or instabilities based on chemorepulsion~\cite{Liebchen2015} 
may help explain the still elusive dynamic clustering observed in experiments~\cite{Theurkauff2012,Palacci2013,Buttinoni2013}.

In all these cases chemotactic instability relies on the ability of weak chemical fluctuations around the uniform density to align microswimmers up (or down) chemical gradients. 
However, under many circumstances microswimmers vary their swimming direction autonomously of chemical cues: this occurs, e.g., for bacteria swimming clockwise close to a glass wall~\cite{Lauga2006}, 
or anti-clockwise near an oil-water interface~\cite{Leonardo2011} (see Electronic Supplementary Information (ESI) \dag for a discussion on parameters).
Synthetic examples of active signalling rotors with self-propulsion (sometimes called 'circle swimmers') may be realised with L-shaped phoretic swimmers~\cite{Kummel2013,Wykes2015}, or with active 
particles with dipole moments~\cite{Baraban2013a,Baraban2013,Palacci2013}
which will track the rotation of applied external fields \cite{Palacci2013,Nguyen2014} and can interact via self-produced phoretic fields \cite{Palacci2013}.   
\\We should expect that, close to uniform states, finite active torques as caused by intrinsic rotations will generally outcompete chemotactic torques which are proportional to the chemical gradient, 
when it comes to determining swimming direction: hence, they might generically lead to a breakdown of the linear KS instability. Does this rule out the possibility of phase separation and patterning in signalling rotors? 
This could have profound consequences for systems like thin (bio)films of chemotactic bacteria~\cite{Berg1990} or actuated phoretic particles.
\\Here, we propose a generic model to study active signalling rotors.
As a key result we find that even weak active rotations suppress the linear KS instability.
However, we identify an instability mechanism which is distinct from the KS mechanism and creates a nonlinear, but deterministic, route to pattern formation. 

For small rotation frequency, this route generates macroscopic phase separation: a fraction of the rotors condenses into a dense phase, separated from a dilute gas by a quasi-stationary interface 
where chemotaxis suppresses rotations and `locks' the average swimming direction onto the upward density gradient.
Away from the interfaces, rotations persist in both phases and lead to dynamic patterns, such as spots or moving stripes and spirals (Fig.~\ref{fig2}h,i) 
This combination of phase separation with pattern formation within both phases represents a novel type of hierarchical structure formation.
When increasing the rotation frequency above a certain threshold, strikingly, the growing clusters do not coarsen any more but form a global pattern of travelling stripes with self-limiting size, 
suggesting that active rotations can be used to control coarsening in suspensions of self-propelled particles (Fig.~\ref{fig2}l).

We describe active particles which self-propel with velocity $v_0$ and rotate with frequencies $\omega_i>0$ at a coarse grained level in two-dimensions.
A smooth $\rho({\bf x},t)$ represents the active particle density and ${\bf p}({\bf x},t)$ is the local average of the unit vector describing the direction of self-propulsion. 
We focus on polarized rotors with identical frequencies, which applies to synthetic colloidal rotors locked to a rotating field, but also to bacterial rotors where short-ranged alignment interactions can synchronize 
the individual rotations locally (see ESI \dag and final paragraph).
Thus, in absence of signalling, ${\bf p}$ rotates with a collective frequency $\omega$. Generally however, our rotors produce signalling molecules with a local rate $k_0\rho$; hence ${\bf p}$ 
also responds, via chemotaxis, to gradients of the resulting chemical field $c({\bf x},t)$, which decays with a rate $k_d$.
This yields a competition of intrinsic active rotations and chemotactic alignment which determines the behaviour of the signalling rotors at large scales. Allowing for finite diffusion of the 
chemical and colloidal density fields with coefficients $D_c$ and $D_\rho$, we describe signalling rotors phenomenologically by:
\begin{figure*}
 \begin{center}
  \includegraphics[width=\textwidth]{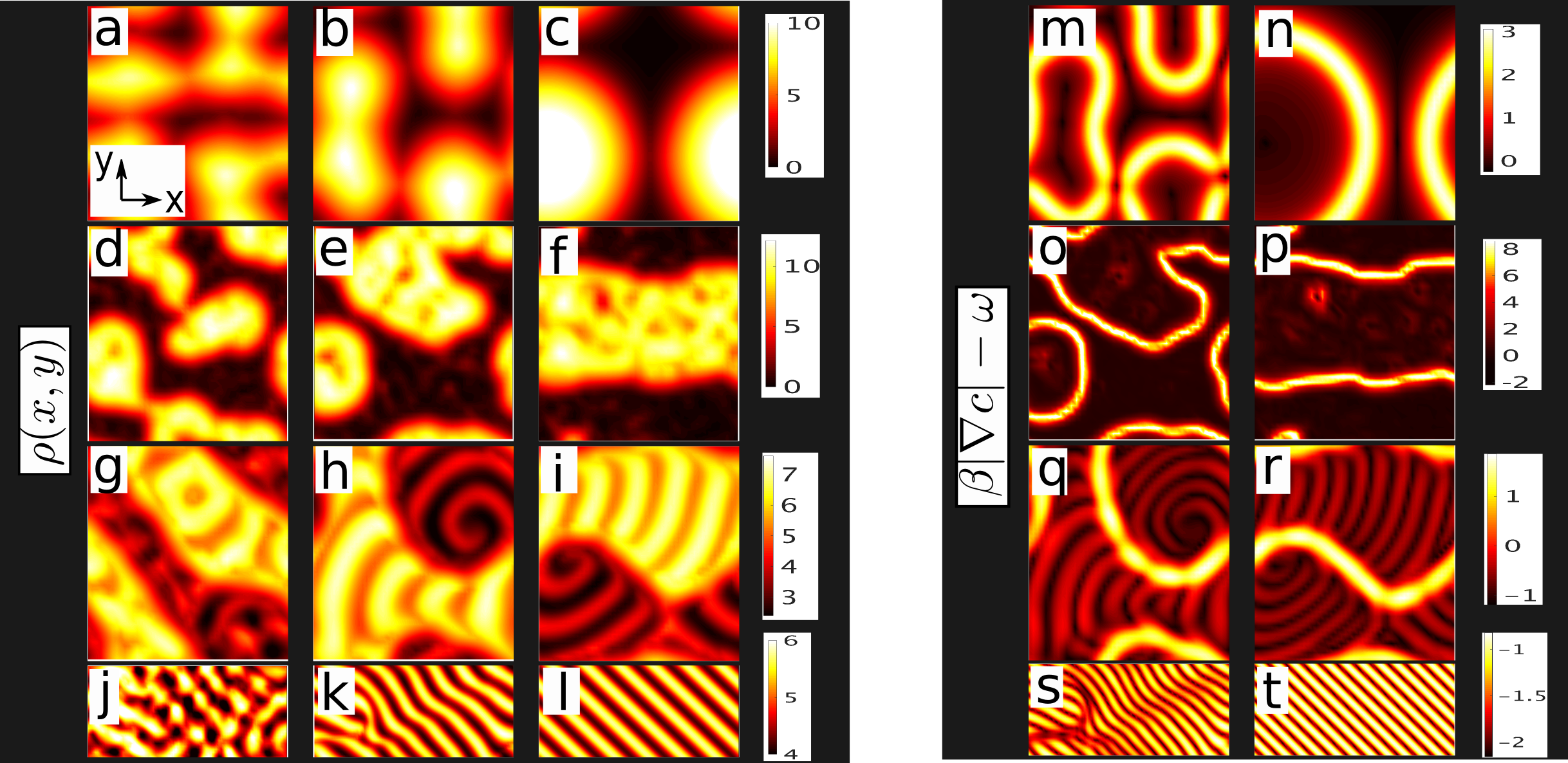}
\caption{\small (a-l): Snapshots of density field $\rho$ for different free rotation frequencies $\omega$ as a function of $x,y$ at different times (increasing from left to right). 
Values for $\omega$ and $\Delta$ are: $\omega=0;\;\Delta=0$ (a-c); $0.25;\;5\times 10^{-3}$ (d-f); $1.2;\;0.024$ (g-i) and $2.2;\;0.044$ (j-l). (m-t): 
snapshots of $\beta|\nabla c|-\omega$ corresponding to the density fields of columns 2 and 3 (m belongs to b; n to c etc). 
Positive values represent locking, negative values rotations. 
Parameters: $\beta=5.0, \rho_0=5.0, v_0=0.5, k_d=k_0=1, \epsilon=10, D_\rho=0.417; K=0.025; ,D_c=1; c_0=(k_0/k_d)\rho_0$; $L_x \times L_y=$ 50$\times$50 in the first and third row, 40$\times$40 in 
the second row and 80$\times$32 in the fourth row; $0.32\sqrt{D_\rho/k_d}$ is the distance between adjacent gridpoints.} 
\label{fig2}
 \end{center}
\end{figure*}
\ea
\dot \rho &=& -v_0 \nabla \cdot(\rho {\bf p}) + D_\rho \nabla^2\rho + K \nabla^2 \rho^3\label{eom1a} \\
\dot \phi &=& \omega + \beta \hat {\bf p} \times \nabla c; \quad {\bf p}=(\cos \phi, \sin \phi)^T \label{eom1b} \\ 
\dot c &=& k_0 \rho - k_d c + D_c \nabla^2 c + \epsilon (c_0-c)^3 \label{eom1c}.
\ee
Here, we used the notation ${\bf a}\times{\bf b}=a_1 b_2-a_2 b_1$ and introduced a chemotactic coupling strength $\beta$, as well as an isotropic short range repulsion among colloids of strength $K>0$ and a cubic reaction term 
with rate coefficient $\epsilon$ leading to the saturation of any unstable mode \cite{note1}.
This phenomenological model can be justified microscopically and is studied here as a minimal description of chemotactic rotors (see ESI \dag), which focuses on the competition between chemotaxis 
and active rotations. 
Its parameter space can be reduced to five dimensions including the (initial) density: only three of these enter the linear dynamics close to the uniform state.
We choose experimentally sensible values for these parameters (ESI \dag) and introduce the dimensionless number $\Delta=\omega v_0/(\rho_0 k_0 \beta)$ as a 
measure of the relative importance of active rotations and chemotaxis.

We now explore the competition of active rotations and chemotaxis by solving
Eqs.~(\ref{eom1a}-\ref{eom1c}) for different $\Delta$ on a square lattice with $L_x \times L_y$ grid points and periodic boundary conditions using finite difference methods.
As an initial state, we choose a small and random perturbation of the spatially uniform and coherently rotating state $(\rho,\phi,c)=(\rho_0,\omega t,(k_0/k_d)\rho_0)$ which solves Eqs.~(\ref{eom1a}-\ref{eom1c}).
In absence of rotations ($\Delta=0$) we observe clusters growing out of the uniform state (Fig.~\ref{fig2}a), which colocalise with chemical density maxima and coarsen at long times (Fig.~\ref{fig2}b), yielding 
one dense macroscopic cluster coexisting with a dilute gas (Fig.~\ref{fig2}c). 
Here, instability of the uniform state is expected due to the positive feedback loop of particle aggregation and chemical production we discussed in the introduction and in \cite{Liebchen2015}. 

For $\Delta=5\times 10^{-3}$ this picture changes dramatically. Now the uniform state persists for a certain duration (see Videos 1,2 in ESI): we call this the initial lag regime. 
Then, almost suddenly, 
fluctuations `awake' and grow to create clusters (Fig.~\ref{fig2}d).
These clusters coarsen into a relatively dense phase separated from a dilute rotor gas (Fig.~\ref{fig2}f) by a slowly-moving interface where rotations are suppressed and particles swim, on average, up 
the chemical gradient (Video 2). 
Away from the interface, all colloids perform a `stop-and-go' rotation, which is associated with dynamic short-lived density dips and peaks (white spots and red dips in yellow region in Figs.~\ref{fig2}e,f). 
Choosing stronger rotations ($\Delta=0.024$) we observe a similar suppression of rotations at the interfaces, but within both phases stripe and spiral patterns form (Fig.~\ref{fig2}g-i and Video 2).

For $\Delta=0.044$ complexity suddenly breaks down: after a long initial lag regime and an intermediate regime where short-lived dynamic clusters continuously emerge and decay, we find a travelling wave 
made of straight parallel stripes (Fig.~\ref{fig2}k,l, Video 4) with self-limiting wavelength. This sudden emergence of a length scale upon increasing $\Delta$ allows control of coarsening via active rotations. 
The length scale of the stripe pattern which decreases as $\omega$ increases and thus can be also controlled. 

\emph{Rotation-locking transition} --
To understand the origin of cluster growth for $\Delta>0$ and the related suppression of rotations at the cluster interfaces, we recast Eq.~(\ref{eom1b}) as
\1 \dot \phi = \omega + \beta|\nabla c| \sin\left(\phi + \delta \right);\quad \delta={\rm arg}(-\partial_x c + \I \partial_y c) \label{adler} \2 
Assuming a time-independent chemical field this can be read, locally, as the Adler equation of synchronization theory which features a transition from (quasi-)periodic rotations ($\beta|\nabla c|<\omega$) 
to phase locking ($\beta|\nabla c| > \omega$), see ESI \dag and \cite{Adler1946,Pikovsky2003,Erneux2010}. 
Although, for our signalling rotors $\nabla c$, hence also $\delta$, evolve in time, in the parameter regime relevant to our numerical simulations, ${\bf p}$ responds fast to changes in $c$, hence we expect 
that the locking transition should still apply\dag. 
Our simulations confirm this expectation and clearly show a transition from `active' rotations to locking at $\beta|\nabla c| \approx \omega$ (Fig.~\ref{fig2}o-r), 
which explains the suppression of rotations at, 
and only at, the interface in Fig.~\ref{fig2}e,f,h,i where chemical gradients are strong enough to generate locking. 
This locking induces a permanent self-advective flux of particles from the dilute phase to the dense phase balancing diffusion in the opposite direction and thus stabilizes dense rotor clusters.

\emph{Locking instability} --
Since the locking mechanism stabilizes clusters, we expect an instability mechanism allowing for the growth of these clusters out of the uniform state. We now resolve this in detail.
Working in the adiabatic limit where ${\bf p}$ responds fast to changes of $c$, Eqs.~(\ref{eom1a}-\ref{eom1c}) simplify significantly.
If $\omega=0$, they resemble the Keller-Segel model (ESI \dag), which leads to phase separation through a long wavelength linear instability -- essentially this is the positive 
feedback described in the Introduction.
In sharp contrast, we show in the ESI \dag that active rotations generically suppress linear instability of the uniform phase in Eqs.~(\ref{eom1a}-\ref{eom1c}).
Physically, this is in line with the intuitive expectation that small chemical gradients caused by fluctuations around the uniform state are too weak to fix the orientations of 
active rotors and cannot promote a linear instability. 
As a result, the KS instability is ineffective for active signalling rotors.

This raises the question why we could still observe cluster growth in our simulations. The answer is that chemotaxis can abduct the rotors into the nonlinear regime before they complete a full rotation, 
and this in turn can lead to a nonlinear instability. To understand the underlying physical mechanism, let us reconsider our coherently rotating uniform initial state, where all rotors are in phase (Fig.~\ref{fig1}c). 
While $\rho,c$ remain approximately uniform in the course of the early-stage dynamics, weak fluctuations of $c$ continuously dephase the orientation field ${\bf p}$ due to chemotaxis (Fig.~\ref{fig1}d). 
Once the rotors are sufficiently out of phase, they can form, temporarily, aster-like `seed' configurations (Fig.~\ref{fig1}e), where all particles in the vicinity of a positive fluctuation of $c$ swim up the chemical gradient. 
This configuration promotes, temporarily, a growth of the fluctuation via the standard KS feedback loop. If this growth generates a chemical gradient surpassing the locking-threshold ($\beta|\nabla c|>\omega$) 
before the seed decays, then ${\bf p}$ remains locked in its vicinity -- leading to a stable cluster as just discussed \cite{note2}.
Hence, we call our nonlinear instability the `locking instability'.
Whether the locking instability is effective or not in practice is a matter of competing timescales, between the 
instantaneous growth rate of chemical gradients and $\omega$-- the former needs to dominate to create locking. Video 5 in the ESI \dag shows the early stage dynamics of $c$ and ${\bf p}$, 
reflecting that $c$ and its gradients can indeed grow beyond the locking threshold on timescales where ${\bf p}$ rotates only slightly.
\begin{figure}
 \begin{center}
  \includegraphics[width=0.48\textwidth]{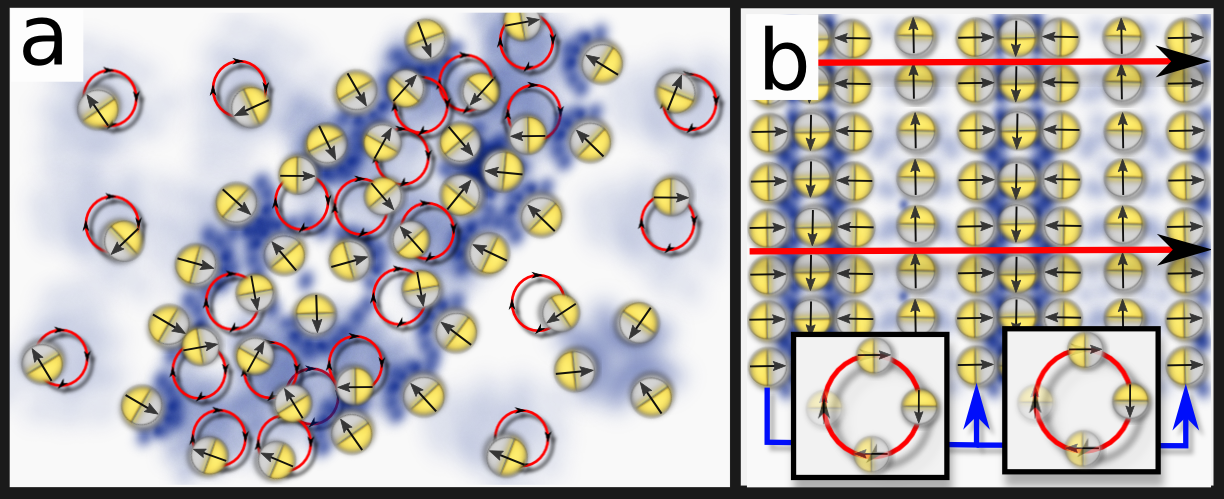}
\caption{\small Schematics of possible asymptotic states (see text) 
a.) Slow rotations: Macrophase separation with coexisting dilute and dense phase of rotors 
separated by an interface of chemically locked rotors.
b.) Moderate rotations: Travelling wave pattern of signalling rotors with synchronized rotational and translational motion of ${\bf p}$.
}
\label{fig3}
 \end{center}
\end{figure}

\emph{Breakdown of locking} --
Locking gets increasingly difficult as $\Delta$ increases and eventually breaks down, quite suddenly, at $\Delta \approx 0.027$ (for parameters as in Fig.~\ref{fig2}); beyond this, 
stable clusters are no 
longer possible.
Specifically, as $\Delta$ increases (i) steeper interfaces are needed to satisfy the locking criterion and (ii) active rotations increasingly dominate the equilibrium between chemotaxis and rotations 
which leads to a steering of the locked ${\bf p}$ away from its `ideal' orientation up the gradient.
This in turn reduces the advective flux up the density gradient leading to shallower interfaces (see ESI \dag and Videos 2,3). 
Unlike for linear (supercritical) instabilities an instability criterion is ill-defined here, and the breakdown of locking depends on an implicit balance between steepness of interfaces and locking orientation of 
${\bf p}$ far away from the uniform state. 
Likewise, when comparing the locking instability to nucleation in systems approaching thermal equilibrium, which is also nonlinear and requires a seed configuration, we find again important differences. 
Crucially, for example, our clusters require steep interfaces to survive, and a 
dense cluster with shallow interfaces would quickly decay; in contrast, for near-equilibrium systems, once a dense seed has nucleated it will grow, however shallow its interfaces may be.

\emph{Active rotations arrest coarsening} -- 
For large $\Delta$, where rotations are too strong to allow locking, we observed a long period where small and short lived clusters dynamically emerge and decay on timescales $\sim 2\pi/\omega$.
These clusters emerge from seeds which promote a growth of $\rho$ and $c$ which is too slow to surpass the locking threshold and hence slows down the rotation of ${\bf p}$ only temporarily.
Strikingly, however, this dynamic clustering does not proceed forever, and eventually some clusters merge (due to diffusion and short ranged repulsions which penalize large interfaces) and synchronise their rotations and translations to form coherent waves (see Video 4 in ESI). 
Specifically, within clusters, the relative phases (frequencies) of adjacent rotors change continously until the system eventually reaches a steady state. Here 
such a steady state is possible in the form of a
stripy pattern (Fig.~\ref{fig2}l), which moves precisely by one spatial period during one rotational cycle of ${\bf p}$. 
In a frame comoving with the pattern, ${\bf p}$ is then fixed (Fig.~\ref{fig3}b) and the pattern, a periodic non-uniform state, is an attractor for the dynamics of the system. Hence, merging and 
approaching this steady state represents a second route allowing short lived clusters to avoid decay.

This process of forming short lived clusters which prevent their decay by merging and moving into a collective direction (spontaneous symmetry breaking) constitutes a second nonlinear instability mechanism of 
the uniform state that is independent of the locking mechanism. In contrast with the latter, this mechanism works at all frequencies, but is comparatively slow since it involves the coordination of several clusters.
Hence this second route to structure formation in signalling rotors is particularly relevant in parameter regimes where the locking mechanism is ineffective, i.e., for large $\omega$, 
but applies similarly to the moving stripes formed within the dense and the dilute phase in Fig.~\ref{fig2}g-i.

As its most remarkable feature and contrasting the locking instability, this route to pattern formation introduces a length scale in steady state. 
Within a large region of parameter space, this length scale increases approximately as $v_0/\omega$. Hence, for synthetic rotors this length scale can be tuned via the frequency of an applied rotating field.
We emphasize that this length scale is determined purely dynamically and cannot be calculated via the standard tools for linear (supercritical) instabilities like amplitude 
equations \cite{Cross1993}.
 
\emph{Non-identical frequencies} --
We finally discuss signalling rotors with non-identical frequencies (e.g., a population of bacteria close to a wall). In absence of alignment effects these rotors quickly dephase and destroy the local polar order. 
However, hydrodynamic \cite{Golestanian2011} or steric alignment interactions can `synchronize' individual rotations. 
Assuming, e.g., short ranged polar alignment interactions with alignment rate
$(g/r_c^2) \sin(\theta_j-\theta_i)$ between particles $i,j$ (polar coordinates)
if their distance $|{\bf r}_i-{\bf r}_j|$ is smaller than a cutoff radius $r_c$ and ignoring chemotaxis (which supports alignment), the local orientation of a rotor follows
\1 \dot \theta_i  = \omega_i + \sqrt{2 D_r} \xi_i(t) + (g/r_c^2) \sum \sin(\theta_j-\theta_i) \label{kur}\2
Here we sum over all $N_c \approx r_c^2 \rho({\bf r}_i)$ particles with distance less than $r_c$ to the $i$th particle and $\xi_i(t)$ represents unit-variance Gaussian white noise.
Remarkably, for dense systems
, Eq.~(\ref{kur}) can be identified (locally) as the Kuramoto model with noise \cite{Acebron2005} which shows a phase transition from the incoherent uniform phase to frequency locking if $g \rho_0$ is a few times larger than 
$D_r$ and the standard deviation $\Delta_\omega$ of the distribution function of the individual frequencies \cite{Acebron2005}.
Thus, within an ensemble of non-identical rotors, 
a strong enough alignment interaction causes a macroscopic fraction of them to rotate in phase with an average frequency, so that our analysis for identical rotors still largely applies. The main caveat is that 
synchronization of the individual rotations takes place locally, so the average rotation may vary in space and time. We thus performed simulations of Eqs.~(\ref{eom1a}-\ref{eom1c}) with additional frequency 
noise $\omega_\xi({\bf x},t)$ in Eq.~(\ref{eom1b}): the results resemble those in Fig.~\ref{fig2} 
as long as $\Delta_\omega<\omega$ and the correlation time of $\omega_\xi$ is short compared to $1/\omega$.
Note, that the ESI \dag material provides an alternative and direct microscopic justification of Eqs.~(\ref{eom1a}-\ref{eom1c}) for rotors with non-identical frequencies.

In conclusion, although even weak active rotations linearly stabilise the uniform phase in ensembles of auto-chemotactic particles, they generate a nonlinear route to structure formation.
This route creates novel patterns including hierarchically organized states combining phase separation and pattern formation. 
It also allows for features which would be impossible to achieve in linear instability scenarios, such as a delayed onset of patterning whose lag time can be programmed via the initial conditions.  
More generally, we showed that rotations provide a versatile new tool to design self-assembly and collective behaviour of active matter, for example to control coarsening.

\emph{Acknowledgements}
BL gratefully acknowledges funding by a Marie Sk{\l}odowska Curie Intra European Fellowship (G.A. no 654908) within Horizon 2020. MEC holds a Royal Society Research Professorship.
Work funded in part by EPSRC EP/J007404/1.


\begin{thebibliography}{34}
\bibitem[Note1()]{note1}
\bibinfo{note}{While we chose a cubic interaction term in 
analogy with the $\phi^4$ theory, a quadratic interaction term $\propto \nabla^2 \rho^2$ with 
appropriately increased coefficient does not change our results. Similarly, 
replacing the cubic reaction term 
by e.g. a quadratic, sign-changing term $\propto (c_0-c)|c_0-c|$ has very little bearing on our results. (A quadratic, non sign-changing term $\propto (c_0-c)^2$ would lead to a blowup of fluctuations rather than to their saturation.)}
\end{thebibliography}
\end{document}


\title{Pattern Formation in Chemically Interacting Active Rotors}

\date{\today}

\begin{center}{\bf \LARGE Electronic Supplementary Information to Pattern Formation in Chemically Interacting Active Rotors}\\\vspace{0.5cm} Benno Liebchen, Michael E. Cates, Davide Marenduzzo
\end{center}

\subsection{Microscopic approach to phenomenological model}
\subsubsection{Rotors with identical frequencies}
Here, we derive coarse grained equations of motion for the rotor density and orientation fields.
Since we are mainly interested in the competition between active rotations and chemotaxis - and in the corresponding physical 
mechanism allowing for structure formation - we use a variety of approximations to achieve a `minimal' rather than a 
rigorous coarse grained description of active rotors which focuses on this competition (and hence only covers a subset of possible phenomena).  
This procedure leads to Eqs.~(\ref{eom1a}-\ref{eom1c}) considered in the main text. 

Consider an ensemble of $N$ `signalling rotors', which self-propel with constant velocity $v_0$ along the directions ${\bf p}_i=(\cos\theta_i,\sin\theta_i)$.
In the uniform state, these directions rotate actively with natural frequencies $\Omega_i$, but generally also respond to chemical gradients, where $\beta_R$ denotes the `chemotactic' coupling strength. 
We further assume (steric) alignment interactions between different rotors which are sufficiently short ranged to allow us to replace their spatial dependence by a pseudopotential (a zero ranged `$\delta$'-interaction).
Denoting the rotational diffusion constant by $D_r$, and by $\xi(t)$ a Gaussian white noise with unit variance, we describe a single chemotactic rotor in 2D via the following Langevin equations
\ea
\dot {\bf r}_i &=& v_0 {\bf p}_i \label{langevinr} \\
\dot \theta_i &=& \Omega_i + \beta_R {\bf p}_i \times \nabla c + G \Sum{j=1}{N} \sin(\theta_i - \theta_j)\delta({\bf r}_i -{\bf r}_j) + \sqrt{2D_r} \xi_i(t)  \label{leqs}
\ee
where we used the notation ${\bf a}\times{\bf b}=a_1 b_2-a_2 b_1$ and $c({\bf x},t)$ is the chemical field produced by the ensemble of `signalling' rotors with rate $k_0$. This field evolves as 
\1 \dot c = k_0 \Sum{i=1}{N}\delta ({\bf r}-{\bf r}_i) - k_d c + D_c \nabla^2 c + \epsilon (c-c_0)^3\2
where $k_d$ is the decay rate. Here, $D_c$ is the chemical diffusion constant and the term proportional to $\epsilon$ prevents unlimited growth of $c$ in case of linear instability. 

We now use It\^os Lemma and follow \cite{Dean1996} to derive coarse grained equations of motion for the combined probability density $f_i({\bf r},\theta)=\delta({\bf r}-{\bf r}_i)\delta(\theta - \theta_i)$.
For rotors with identical frequencies $\Omega_i \rightarrow \Omega$ we find for $f({\bf r},\theta)=\Sum{i=1}{N}f_i({\bf r},\theta)$
\1 \dot f =-v_0 {\bf p}\cdot \nabla f - G \partial_\theta \int \de {\bf r}'\de {\bf \theta}'f({\bf r}',\theta') 
F(\theta-\theta',{\bf r}-{\bf r}') f({\bf r},\theta)+D_r \partial^2_\theta f-\beta_R|\nabla c|\partial_\theta [f \sin(\theta+\delta)]-\partial_\theta \Omega f -\partial_\theta \sqrt{2D_r f}\xi(t) \label{dean}\2
Here $\delta={\rm atan2}(\partial_y c,-\partial_x c)$, where ${\rm atan2}(y,x)$ is the (principle value of) the argument function ${\rm arg}(x+\I y)$
and $-\partial_\theta \sqrt{2D_r f}\xi$ describes multiplicative noise, which we neglect in the following because we are interested in the mean field phenomenology. 

From here, we follow \cite{Bertin2009} and expand $f$ in a Fourier series $f(r,\theta)=\Sum{k=-\infty}{\infty}f_k({\bf r}){\rm e}^{-\I k \theta}$ with $f_k({\bf r})=\int f({\bf r},\theta) {\rm e}^{\I k \theta}\de \theta$ and identify $f_0 \rightarrow \rho({\bf r},t)$ and $\left({\rm Re}f_1,{\rm Im} f_1\right) \rightarrow {\bf w}$.
\footnote{This corresponds to a `continuum' approximation which is, due to the interaction term, only sensible in systems that are dense enough such that the interaction can be averaged over many neighbours.}
Straightforward algebra leads to the following equation for the Fourier coefficients: 
\1  
\dot f_k = 
-\4{v_0}{2}\left[(\partial_x-\I \partial_y)f_{k+1} + (\partial_x+\I\partial_y)f_{k-1}\right]
-k^2 D_r f_k +\4{\I G k}{2\pi} \Sum{m}{}f_{k-m} F_{-m}f_{m}
+\4{\beta_R |\nabla c| k}{2} \left(f_{k+1}{\rm e}^{\I\delta} - f_{k-1} {\rm e}^{-\I \delta}\right) + \I k \Omega f_k \label{fpn}
\2
For $k=0$ we quickly find $\dot \rho = - v_0 \nabla \cdot {\bf w}$. 
To achieve a closed equation for $f_1$ we neglect $f_k$ with $k\geq 3$ and assume that $f_2$ is fast, i.e. we set $\dot f_2 \rightarrow 0$ (compare \cite{Bertin2009,Farrell2012}).
This yields a closed equation of motion for $f_1$ and hence for ${\bf w}$. This latter equation could in principle be solved numerically, but is rather involved and hardly allows us to analyse the interplay between chemotaxis and active rotations. 
To highlight this interplay within a `minimal model' we apply a second layer of approximations and directly neglect all contributions which are both nonlinear and involve gradient terms (such terms would therefore contribute to the nonlinear saturation of unstable short wavelength modes)\footnote{These neglected terms should be important at least deep in the nonlinear regime and could be the subject of further investigations.}. This procedure leads to
\ea
\dot \rho &=& - v_0 \nabla \cdot {\bf w} \label{rf}\\
\dot {\bf w} &=& \left(\4{G \rho}{2}-D_r\right){\bf w} + \Omega {\bf w}_\perp- \4{v_0}{2}\nabla \rho + \4{\beta_R \rho}{2}\nabla c + \4{v_0^2 D_r}{b}\nabla^2 {\bf w} + \left(\4{v_0^2 \Omega}{2b}\right)\nabla^2 {\bf w}_\perp
-\left(\4{2G^2 D_r}{b}\right)|{\bf w}^2|{\bf w} - \4{G^2 \Omega}{b}|{\bf w}|^2 {\bf w}_\perp \label{wf}\\
\dot c &=& k_0 \rho - k_d c  + D_c \nabla^2 c + \epsilon(c-c_0)^3 \label{cf}
\ee
where ${\bf w}_\perp\equiv(-w_y,w_x)$ and $b=16D_r^2 + 4\Omega^2$.
Two solutions of these equations are: (i) the uniform unpolarized state $(\rho,{\bf w},c)=(\rho_0,{\bf 0},\rho_0 k_0/k_d)$, and (ii) a polarized and coherently rotating state with uniform density $(\rho,w,\phi,c)=\left(\rho_0, w^\ast,\Omega^\ast t, k_0 \rho_0/k_d \right)$ with $w^\ast=\sqrt{[G\rho_0-2Dr](\Omega^2/(G^2D_r)+4D_r/G^2)}$, $\Omega^\ast=\Omega(3/2-G\rho_0/(4D_r))$.
Here, $\phi$ is defined via ${\bf w}=w {\bf p}$ with ${\bf p}=(\cos\phi,\sin\phi)^T$ representing the average (collective) self-propulsion direction.
If $G \rho_0 < 2 D_r$ the unpolarized state is stable. It becomes unstable in favour of the coherently rotating state when alignment interactions are strong enough to suppress dephasing by rotational noise ($G \rho > 2 D_r$). At the onset of polarization this state rotates with a frequency $\Omega$ but slows down as more and more particles align.

In the main text, we focus on the regime of sufficiently strong alignment interactions and hence consider this coherently rotating state as the relevant uniform state whose dynamics we explore in presence of chemotaxis. 
Also, if self-propulsion is not too strong, the term $(-v_0/2)\nabla \rho$ mainly reduces the chemotactic coupling in our simulations where we typically have $c\sim \rho$
and $\beta_R\rho_0 \gg v_0$. Hence we also omit this term for simplicity 
as well as $\nabla^2$ terms in Eq.~(\ref{wf}), which are not important at long wavelength or for almost uniform ${\bf w}$.
We now define the polarization ${\bf P}$ which measures the degree of local alignment (per particle), as usual, via ${\bf w}=\rho {\bf P}=\rho |{\bf P}| {\bf p}$.
Since we are mainly interested in the competition of the chemical alignment and active rotations, we do not describe spatial modulations of the `average' polarization but assume it as constant; $|{\bf P}|=1$ for simplicity (smaller positive values of $|{\bf P}|$ do not significantly alter our numerical results).
Assuming also $\omega:=\Omega-\frac{G \Omega}{b}{w}^2\approx \Omega-\frac{G \Omega}{b}{\rho_0^2}$ in Eq.~(\ref{wf}) and defining $\beta:=\4{\beta_R}{2}$, after projecting Eq.~(\ref{wf}) onto $\hat {\bf p}_\perp=(-\sin \phi,\cos \phi) $ our minimal description of chemotactic rotors reads
\ea
\dot \rho &=& -v_0 \nabla\cdot(\rho {\bf p}) + D_\rho \nabla^2 \rho + K \nabla^2 \rho^3 \label{rhosi}\\
\dot \phi &=& \omega + \beta {\bf p} \times \nabla c \label{adlersi} \\
\dot c &=& k_0 \rho - k_d c  + D_c \nabla^2 c + \epsilon(c-c_0)^3 \label{csi}
\ee
In Eq.~(\ref{rhosi}), we added a phenomenological term describing isotropic short ranged repulsions among colloids ($K \nabla^2 \rho^3$), whose main effect is to prevent strong gradients on too small scales in our simulations. 
While we chose here a cubic term for convenience (retaining symmetry under $\rho\rightarrow -\rho$), replacing this by a quadratic term, $\nabla^2 \rho^2$ leads, according to our simulations, to an almost identical phenomenology when modifying the coefficient $K$ appropriately.
Given the complexity of the system under consideration, our description in Eqs.~(\ref{rhosi}-\ref{csi}) is far from complete, but it highlights the competition between chemotactic alignment and active rotations, allowing us to focus on the physical mechanism underlying structure formation in signalling rotors. Our approach can be straightforwardly extended to derive a more precise but rather complex coarse grained description of (chemotactic) active rotors.

\subsubsection{Generalization to non-identical rotation frequencies}
Here, we generalize the above approach to rotors with non-identical frequencies $\Omega_i$, i.e. we replace $\Omega \rightarrow \Omega_i$ in Eq.~(\ref{leqs}) 
and ask to which extend this alters Eqs.~(\ref{rhosi}-\ref{csi}).
Accordingly, this paragraph can be seen as an alternative to the paragraph `non-identical frequencies' in the main text.
\\The basic idea is to replace $f_i\Omega_i$ by its mean plus a typical fluctuation
\1 \Omega_i f_i \longrightarrow \bar \Omega f + \sqrt{\Delta_\Omega f} \eta \label{drepl}\2
where $\eta$ describes Gaussian random numbers with zero mean and unit variance $\langle\eta({\bf r},\theta,t)\eta({\bf r}'\theta',t)\rangle= \delta({\bf r}-{\bf r}')\delta(\theta-\theta')$ and $\bar \Omega$;
$\Delta_\Omega$ are defined by 
\1 \langle{\Omega_i \Omega_j}\rangle ={\bar \Omega}^2 + \Delta_\Omega \delta_{ij} \label{corr} \2
This replacement of course changes, in general, the dynamics of $f_i$ but not the statistical properties of a large rotor ensemble, since the 
distributions $\{\Omega_i f_i\}$ and $\{\bar \Omega f + \sqrt{\Delta_\Omega f} \eta\}$ have identical statistical properties. (This statement holds true at each point in time, 
independent of time correlations of $\eta$).
\\Physically, we should understand Eq.~(\ref{drepl}) as a local replacement and interpret 
the above averages as mesoscopic ones over all particles within the interaction range around a given point ${\bf x}$ rather than a global average over the whole rotor ensemble.  
Accordingly, we allow $\bar \Omega$ and $\eta$ to fluctuate in space and time.
We now define a field $\omega({\bf x},t)$ representing the deviation from the global (time-independent) average rotation frequency
$\bar \Omega_0=\Sum{i=1}{N}\Omega_i/N$. 
\\In the following we do not worry about the specific form of $\omega({\bf x},t)$ but assume it to be random, with a correlation time
on the order of the `mixing time', i.e. the time 
after which a given set of rotors in
one `interaction domain' is replaced by another one. In non-synchronized states, we expect that the mixing time is on the order of the time a particle needs to traverse a distance given by the range of the alignment interactions. 
As this timescale is, for the assumed short ranged alignment interactions, short compared to all other relevant timescales in the system ($1/k_d, 1/k_0, 1/\omega$) we 
allow it to tend to zero for simplicity.
Hence we assume $\omega$, and analogously also $\eta$, to represent spatiotemporal white noise.
Conversely, in the synchronized regime, rotors can move together for some time and hence we expect substantial time-correlations. 
As these correlations should become relevant only after the onset of synchronization we do not need to care about them since our aim is to understand where synchronization sets in.

We now expand $\sqrt{f}$ for modest deviations from isotropy as follows
\1 
\sqrt{f}=\sqrt{\Sum{k=-\infty}{\infty} f_k {\rm exp}[-\I k \theta]} \approx \sqrt{f_0}/2+\4{1}{2\sqrt{f_0}}\Sum{k=-\infty}{\infty} f_k {\rm e}^{-\I k \theta}.\label{tex}
\2
Using this approximation, the only modifications of our mean field results Eqs.~(\ref{rf}-\ref{csi}) due to non-identical rotor frequencies
correspond to replacing 
\ea \bar \Omega &\rightarrow& \bar \Omega_0 + \omega({{\bf x},t}) \\ 
D_r &\rightarrow& D_r + \sqrt{\4{\Delta_\Omega}{4\rho_0}} \label{Drnif}
\ee
in these equations.
Consequently, for non-identical rotors the unpolarized state becomes unstable for $G\rho_0 > 2D_r +  \sqrt{\4{\Delta_\Omega}{4\rho_0}}$.
This is very similar to the result of the Kuramoto model (main text) which assumes a Lorentzian distribution for the frequencies, 
and justifies the assumption of locally coherent rotations underlying model (\ref{rhosi}-\ref{csi}).


\subsection{Nondimensionalisation and relation to experimental parameters}
The parameter space of Eqs.~(\ref{rhosi}-\ref{csi}) can be reduced to four dimensions (plus an effective density which is fixed by the initial state) by introducing the dimensionless quantities $\tilde x = \sqrt{k_d/D_\rho}x$, $\tilde t = k_d t$. 
Defining $\tilde \rho = \rho k_0 \beta/(k_d v_0)$, $\tilde c = c \beta/v_0$ and $\tilde {\bf p}=v_0 {\bf p}/\sqrt{D_\rho k_d}$ we obtain (now omitting tildes)
\ea
\dot \rho &=& -\nabla\cdot (\rho {\bf p}) + \nabla^2 \rho + \kappa \nabla^2 \rho^3 \label{eomru1} \\
\dot \phi &=& \Omega + {\bf p} \times \nabla c  \label{eomru2} \\ 
\dot c &=& \rho - c + \mathcal{D}_c \nabla^2 c + \varepsilon (c_0 - c)^3 \label{eomru3}
\ee
Here, $\Omega=\omega/k_d$ and $\mathcal{D}_c=D_c/D_\rho$ determine the linear behaviour together with the effective density $\rho_0$ 
which is conserved in the course of the dynamics,
while $\kappa=K k_d^2 v_0^2/(D_\rho k_0^2 \beta^2)$ and $\varepsilon=\epsilon v_0^2/(k_d \beta^2)$ control non-linear saturation effects (besides the $\kappa$ term producing
some contribution to the colloidal diffusion term). %
As a key control parameter we identify $\Delta:=\Omega/\rho_0 = v_0 \omega/(\beta k_0 \rho_0)$ which measures (for given self-propulsion speed) 
the relative importance of active rotations and chemotaxis.

Promising candidates to physically realize our predictions are auto-chemotactic strains of E.coli which rotate naturally close to a wall or interface. 
Here, one could measure the initial time lag followed by a delayed onset of clustering as an indication of the nonlinear locking instability. 
In addition, one could also measure the suppression of rotations which arises exclusively at the interface between dense and dilute regions for comparatively low effective rotation frequencies (see below) and the sudden arrest of coarsening when for larger rotation frequencies. 

The parameter values used in this work (Fig.~\ref{fig2}) correspond to typical experimental values; the most 
relevant dimensionless control parameters $\Omega$, $\tilde \rho$ and $\mathcal{D}_c$ translate as follows. (i) 
The choice $D_c/D_\rho = 1$ approximately matches with measurements of $D_\rho \sim 2\times 10^{-6}-1.5\times 10^{-5} {\rm cm}^2/{\rm s}$ 
and $D_c \sim 10^{-5} {\rm cm}^2/{\rm s}$~\cite{Tindall2008,Ford1991,Berg1990,Murray2003}. (ii) Our parameter choice in Fig.~\ref{fig2} 
corresponds to $\rho_0 \beta k_0/k_d v_0 =50$ which can be matched for typical `chemotactic sensitivities' of $\chi \sim (1.5-75) \times 10^{-5} {\rm cm}^2/{\rm s}$ \cite{Tindall2008,Ford1991} 
with $\chi \sim \beta v_0/(2D_r)$ when assuming $D_r \sim 0.1-1/{\rm s}$ and rates $k_d \sim 0.09/{\rm s}$ and $k_0\sim 0.6/{\rm s}$ typical for the chemoattractant of 
Dicty (cAMP~\cite{Calovi2010,Martiel1987}, rates for E.coli are unknown \cite{Murray2003}) when using systems with $>5\times 10^3$ bacteria per ${\rm mm}^2$ 
or area fractions $>0.1$. (iii) The parameter $\Omega=\omega/k_d \sim 2$ for typical swimming radii of $50\mu {\rm m}$ \cite{Lauga2006} and swimming speeds of $10\mu {\rm m}/{\rm s}$ 
which is close to the value at which we observed the transition from a large cluster to a stripe pattern; see Figs.~\ref{fig2}i and l. Faster swimming leads to 
larger values of $\Omega$ and could allow to  observe the decrease of the wavelength of our travelling wave pattern. Smaller values of $\Omega$ 
corresponding to parameters in Figs.~\ref{fig2}d-f would require slower swimming velocities or faster decay (consumption) of the E.coli chemoattractant aspartate than for cAMP. 
Alternatively, by working with magnetotactic bacteria which have a permanent dipole moment or active bimetallic colloids \cite{Palacci2013}, slow rotations could be easily generated by exposing 
these particles to an external rotating magnetic field.

\subsection{Reduced models and linear stability analysis}
We now derive a reduced model for the rotor distribution $\rho({\bf r},t)$ by assuming that ${\bf p}$ is a fast variable, which is a good approximation for the parameter regime considered in the main text. This allows us to adiabatically eliminate the orientation equation. 
First, we demonstrate that in absence of active rotations and alignment interactions $\Omega=G=0$, Eqs.~(\ref{rf}-\ref{cf}) reduce to the well-known Keller-Segel model \cite{Keller1970,Keller1971} describing chemotaxis of signalling microorganisms, and are therefore consistent with previous studies of active colloids based on this model \cite{Theurkauff2012,Meyer2014,Liebchen2015}.
In particular, the Keller-Segel model for chemoattractive particles ($\beta>0$) allows for cluster growth typically proceeding to phase separation. Here, as we shall see, local alignment interactions ($G>0$) will support this instability in the sense that they allow for cluster growth already at very low particle densities. 
Second, we will derive a corresponding reduced model in presence of active rotations, starting with Eqs.~(\ref{eom1a}-\ref{eom1c}) in the main text.
This model shows that active rotations suppress linear instability in its complete parameter space, even if strong alignment interactions are present. 

\subsubsection{Keller-Segel model} 
Starting from Eq.~(\ref{wf}) and adiabatically eliminating $\dot {\bf w}\rightarrow {\bf 0}$ as well as neglecting gradient terms of order $\nabla^2$, with 
$\Omega=G=0$ we find ${\bf w}=-v_0 \nabla \rho/(2D_r) + \beta \rho \nabla c/{2D_r}$. Plugging this into Eq.~(\ref{rf}) leads together with Eq.~(\ref{cf}) to the following model which resembles the Keller-Segel model of chemotaxis (compare \cite{Keller1970,Keller1971,Theurkauff2012,Meyer2014,Liebchen2015}) 
\ea
\dot \rho &=& - \4{v_0 \beta}{2D_r} \nabla \cdot (\rho \nabla c) + D \nabla^2 \rho \\
\dot c &=& k_0 \rho - k_d c + D_c \nabla^2 c + \epsilon (c-c_0)^3
\ee
Here $D=v_0^2/(2D_r)$ is the effective `active' diffusion constant.

Linearizing around the uniform solution $(\rho,c)=(\rho_0,k_0 \rho_0/k_d)$ and using $\rho'=\rho - \rho_0, c'=c-c_0$ leads to:
\ea \dot \rho' &=& - B \nabla^2 c' + D\nabla^2 \rho' \nonumber \\
\dot c' &=& k_0 \rho' - k_d c' + D_c \nabla^2 c' \label{links}
\ee
where $B=\4{v_0 \beta \rho_0}{2D_r}$.  We now test the stability of these equations with respect to plane wave perturbations. 
Fourier transforming (\ref{links}) and evaluating the respective linear stability problem we find a long wavelength instability for $B k_0>D k_d$ with instability band $q^2 < (Bk_0 - D k_d)/(D D_c)$. 
This instability is based on the positive feedback loop explained in our introduction (see main text) and hinges on aligning particles up the chemical gradient.
Accordingly, it is no surprise that local alignment interactions only help destabilize the uniform state and allow for cluster growth already at very low effective densities.
For a later comparison with the case of active rotations it is instructive to further reduce model (\ref{links}). 
Assuming that $c'$ is fast compared to the conserved colloidal field after adiabatic elimination $c'\rightarrow 0$ and neglecting $\mathcal{D}_c \nabla^2 c'$ Eq.~(\ref{links}) simplifies to 
\1 \dot \rho' = [D - (k_0/k_d)B]\nabla^2 \rho' \label{onevarks}\2
This is a closed model for the particle density close to the uniform state where chemotaxis directly competes with the effective diffusion. This equation reproduces the correct instability criterion, but leads to a short wavelength divergence, normally prevented by chemical diffusion.

\subsubsection{Adiabatic solution of the Adler equation}
To derive a reduced model for active rotors ($\Omega>0$) we need to solve Eq.~(\ref{eomru2}) explicitly. Rewriting Eq.~(\ref{eomru2}) leads to the Adler equation $\dot \phi = \Omega + |\nabla c|\sin(\phi+\delta)$, here with $\delta={\rm atan2}(\partial_y c, -\partial_x c)$. We now solve this equation in the adiabatic limit of fast response of $\phi$ to changes in the chemical field.
Since in presence of active rotations the collective propulsion direction ${\bf p}$ as determined by $\phi$ does not simply follow the chemical field but is determined by a competition between chemotaxis and active rotations, we may not simply set $\dot \phi\rightarrow 0$, but need to integrate Eq.~(\ref{adler}), MT for a (quasi-)stationary chemical field to describe fast orientational response appropriately. 
In this adiabatic approximation we find (compare also \cite{Adler1946})
\1 \phi_{\rm ad} = -{\rm atan2}(\partial_y c,-\partial_x c) + 2\arctan\left[\4{-|\nabla c| + \sqrt{\Omega^2 - |\nabla c|^2} \tan\left(\sqrt{\Omega^2 - |\nabla c|^2}t/2 + \theta_0 \right)}{\Omega}  \right] \label{phiad} \2 
Here $\theta_0({\bf x})$ is an integration constant that is fixed by the initial orientations of the rotors. 
For $|\nabla c|<\Omega$ (\ref{phiad}) describes an anharmonic and anisotropic periodic oscillation. As the chemical gradient increases, the rotation frequency decreases as $\sqrt{\Omega^2- |\nabla c|^2}$, i.e. ${\bf p}$ rotates slower where chemical gradients are strong.

At $|\nabla c|=\Omega$ the colloids cease to rotate (on average), i.e. we have a transition from rotations to locking. Here, we have an equilibrium between chemotaxis and rotations, where rotations steer the director field away from the `optimal' swimming direction up the chemical gradient (which is approached for $|\nabla c|\rightarrow \infty$), and towards a direction perpendicular to it.
From Eq.~(\ref{phiad}) it is straightforward to derive the distance from the `optimal' angle as
\1 \Delta:=\phi_{\rm ad}(\Omega)-\phi_{\rm ad}(\Omega\rightarrow 0)=2\arctan\left[\4{\gamma}{1+\sqrt{1-\gamma^2}}\right] \label{lockeddir}\2 
where we used the notation $\gamma:=\Omega/|\nabla c|$. For $\Omega < 1/2$ we can approximate $\Delta \approx \gamma$ and conversely, for $\Omega$ close to $|\nabla c|$, we have $\Delta \approx \pi/2 - \sqrt{2(\Omega/|\nabla c|-1)}$.
In other words, the advective particle flux up density (chemical) gradients ($\propto \cos\Delta$) decreases slowly with increasing $\Omega$ when $\Omega$ is small, but decreases faster and faster for large $\Omega$.

\subsubsection{Linear stability of active rotors}
We now use Eq.~(\ref{phiad}) to formulate a reduced model for the density of rotors whose orientation responds quickly to changes in the chemical field.
For a quasi-instantaneous orientational dynamics Eqs.~(\ref{eomru1},\ref{eomru3}) lead to 
\ea \dot \rho &=& \nabla^2 \rho + \kappa\nabla^2 \rho^3- \nabla \cdot \left(\rho {\bf p}_{\rm ad}(t) \right)  \nonumber \\
\dot c &=& \rho - c + \mathcal{D}_c \nabla^2 c \label{KS1}
\ee
where ${\bf p}_{\rm ad}=\left(\cos(\phi_{\rm ad}),\sin(\phi_{\rm ad})\right)^T$ and $\phi_{\rm ad}$ is given by Eq.~(\ref{phiad}).
To understand the linear stability of the uniform solution $(\rho,c)=(\rho_0,\rho_0)$ of these equations with $\theta({\bf x})={\rm const}$, we choose a coordinate system with x-axis parallel to ${\bf p}(t=0)$, i.e. $\theta=\arctan\left[(|\nabla c|+\Omega \tan(\delta/2))/\sqrt{\Omega^2-|\nabla c|^2}\right]$ and  
expand ${\bf p}_{\rm ad}$ in $\partial_x c,\partial_y c$:
\1
{\bf p}_{\rm ad} = \left(\begin{matrix} \cos\Omega t\\ \sin\Omega t \end{matrix}\right)+ \4{1-\cos\Omega t}{\Omega}\left(\begin{matrix} \sin\Omega t\\ -\cos\Omega t \end{matrix}\right)\partial_x c
+ \4{\sin\Omega t}{\Omega}\left(\begin{matrix} -\sin\Omega t\\ \cos \Omega t \end{matrix}\right)\partial_y c \label{dcexp}
\2
Hence, we obtain:
\ea \dot \rho' &=& (1+3\rho_0^2 \kappa) \nabla^2 \rho'- \left(\begin{matrix} \cos\Omega t\\ \sin\Omega t \end{matrix}\right)\nabla \rho' 
+ \4{\sin^2 \Omega t + \cos\Omega t - \cos^2 \Omega t}{\Omega}\partial_x \partial_y c - \4{\sin\Omega t}{\Omega}\left[1-\cos\Omega t\right]\partial^2_x c - \4{\cos\Omega t \sin \Omega t}{\Omega}\partial^2_y c
\nonumber \\
\dot c' &=& \rho' - c' + \mathcal{D}_c\nabla^2 c' \label{KSols} \ee
Remarkably, the active rotations enter the linearized equations as explicitly time-dependent `driving' terms.
To understand how this leads to linear stability of the uniform state, we first consider the instructive case of fast chemical dynamics $\dot c \rightarrow 0$, before generalizing our approach in the next paragraph. 
Neglecting also chemical diffusivity ($\mathcal{D}_c \rightarrow 0$) which only helps stabilizing the uniform state, we have $c' = \rho'$.
Now using a plane wave Ansatz $\rho'=r(t){\rm exp}[-\I {\bf q}\cdot {\bf x}]$ in Eqs.~(\ref{KSols}), Floquet theory predicts $r(t) =r(0) \exp[\mu t] g(t)$ \cite{Chicone1999}, where $g(t)$ is some $2\pi/\Omega$-periodic function.
Here, $\mu$ is called the Floquet exponent and serves as a linear stability parameter; if it is positive $r(t)$ oscillates with growing amplitude, but for $\mu<0$  small density fluctuations around the uniform state decay. According to the Liouville formula in Floquet theory \cite{Chicone1999}, here $\mu$ is given by the time-average of the right hand side of Eqs.~(\ref{KSols}). Thus, we have $\mu = -(1+3\rho_0^3 \kappa) {\bf q}^2$ meaning that active rotations {\it suppress} the chemotactic instability, which occurs in their absence (Eq.~\ref{onevarks}).
More specifically, the time-dependent driving terms in Eq.~(\ref{KSols}) effectively suppress the impact of self-advection on the dynamics: in the absence of rotations this leads to a spinodal instability (see Eq.~\ref{onevarks}).
This confirms the intuitive argument given in the introduction of the main text: close to the uniform state, chemical gradients are too weak to align the swimming directions of our rotors and so rotations dominate chemotaxis. 
As we discuss in the main text, this picture breaks down away from the uniform state and can give rise to a nonlinear instability allowing for structure formation with chemically interacting rotors despite the fact that the uniform state is now linearly stable. 
 
\subsubsection{Generalized linear stability analysis: Floquet-Magnus expansion}
Finally, we generalize the linear stability analysis of the previous paragraph 
to cases where we may not assume that $c$ is a fast variable and have to account for corresponding delay effects. 
We first plug the Ansatz
$\rho'=r(t){\rm exp}[-\I {\bf q}\cdot {\bf x}]$,
$c'=C(t){\rm exp}[-\I {\bf q}\cdot {\bf x}]$ into Eqs.~(\ref{KSols}), which leads to
\ea
\left(\begin{matrix} \dot r \\ \dot C \end{matrix}\right)=
A \left(\begin{matrix} r \\ C \end{matrix}\right) \label{floqp}
\ee
where the $2\pi/\omega$-periodic matrix $A=(a)_{ij}$ has the following components 
\ea
a_{11} &=& -{\bf q}^2 (1+3\rho_0^2 \kappa)-\I [q_x \cos\Omega t+q_y \sin\Omega t]\\
a_{12} &= &  \4{\rho_0}{\Omega}\left[q_x q_y(\cos^2\Omega t -\sin^2\Omega t-\cos \Omega t) + (q_y^2 - q_x^2)\cos\Omega t\sin\Omega t + q_x^2 \sin\Omega t\right] \\
a_{21} &= & 1 \\
a_{22} &=& -\mathcal{D}_c {\bf q}^2
\ee
Denoting the fundamental matrix solution to Eq.~(\ref{floqp}) as $X$, we have $\dot X(t)=A(t)X(t)$, where $X(t)=P(t){\rm exp}[ t B]$
with $P(t+2\pi/\omega)=P(t)$, according to the Floquet theorem \cite{Chicone1999}. The eigenvalues of $B$ are the Floquet exponents $\mu_{1,2}$ which determine
the linear stability of the present problem. 
From here, we follow \cite{Blanes2009} to perform a Floquet-Magnus expansion, i.e. we expand $P,B$
\ea P &=& {\rm exp}[\Lambda]; \quad \Lambda = \Sum{k=1}{\infty}\Lambda_k; \quad \Lambda_k(t+T)=\Lambda_k(t) \\
B&=&\Sum{k=1}{\infty} B_k
\ee
and determine $B$ up to order $k=2$.
To first order, we find 
\1  \mu_1 = -(1+3\rho_0^2 \kappa){\bf q}^2; \quad \mu_2 = -1 - \mathcal{D}_c {\bf q}^2 \2
which suggests linear stability for all values of $\Omega$.
The general result at second order ($k=2$) is more involved and its physical meaning is
not immediate; thus we perform an asymptotic expansion up to second order in $\nu=1/\Omega$ which yields
\1
\mu_{1,2} = -\4{1}{2}\left[1+ (1+3\rho_0^2 \kappa + \mathcal{D}_c) {\bf q}^2 + \left| 1+ (\mathcal{D}_c - 1 - 3\rho_0^2 \kappa){\bf q}^2\right|\right] \pm 
\I \4{2q_x {\bf q}^2 \rho_0}{\Omega^2 \left| 1+ (\mathcal{D}_c - 1 - 3\rho_0^2 \kappa){\bf q}^2\right|} 
\2
The real parts of $\mu_{1,2}$ (Lyapunov exponents) represent the average growth rate of small perturbations of the uniform state. 
Thus, even if $c$ is not a fast variable, at least for large $\Omega$ active rotations suppress linear stability. 
Note that convergence of the Floquet-Magnus expansion can be shown here for large $\Omega$ when choosing the 2-norm of $A$ to satisfy the
convergence criterion in \cite{Blanes2009}. 
\\To generalize our approach further, for small $\Omega$ we calculated the Floquet exponents $\mu_{1,2}$ numerically for different portions of the parameter space.
Here, we found that $\mu_{1,2}$ have negative real parts (linear stability of the uniform state)
in most regions of the parameter regime, even for relatively slow driving.
However, interestingly, 
they can become positive if 
the effective $\rho_0$ is large enough and $\kappa,\mathcal{D}_c,\Omega$ are sufficiently small. This demonstrates that besides the nonlinear instability mechanism we discussed in the main text, also 
delay effects in the response of $c$ to changes in $\rho$
could provide a route to structure formation in slowly rotating chemically interacting particles. Although this linear route to structure formation applies only to a restricted subset of the parameter space, it 
could be an interesting topic for further investigations.

\subsection{Videos} 
Video 1 shows the dynamics of signalling rotors at short and intermediate times for $\Omega=0$ (this corresponds to Fig.~\ref{fig2}a-c, main text); $\Omega =1.2$ (Fig.~\ref{fig2}g-i); $\Omega =2.2$ (Fig.~\ref{fig2}j-l) 
and $\Omega=7$. Initial states are identical in all four cases. 
The two latter cases ($\Omega=2.2$ and $\Omega=7$) lead to travelling wave patterns on timescales beyond the ones shown in the video. Other parameters are chosen as in Fig.~\ref{fig2}.
Videos 2,3,4 show the rotor dynamics ($\rho$ and ${\bf p}$) on comparatively long timescales for 
$\Omega=0.25$ (This case corresponds to Fig.~\ref{fig2}d-f, but for larger system size), $\Omega=1.2$ (g-i) and $\Omega=2.2$ (j-l), respectively.
Video 5 shows $c$ and ${\bf p}$ for parameters as in Fig.~\ref{fig2}d-f and comparatively short timescales.